# DETECTING AI-ASSISTED CHEATING IN ONLINE EXAMS THROUGH BEHAVIOR ANALYTICS

Gökhan Akçapınar

*Department of Computer Education and Instructional Technology*
*Hacettepe University Beytepe, 06800, Ankara, Turkey*

## ABSTRACT

AI-assisted cheating has emerged as a significant threat in the context of online exams. Advanced browser extensions now enable large language models (LLMs) to answer questions presented in online exams within seconds, thereby compromising the security of these assessments. In this study, the behaviors of students (N = 52) on an online exam platform during a proctored, face-to-face exam were analyzed using clustering methods, with the aim of identifying groups of students exhibiting suspicious behavior potentially associated with cheating. Additionally, students in different clusters were compared in terms of their exam scores. Suspicious exam behaviors in this study were defined as selecting text within the question area, right-clicking, and losing focus on the exam page. The total frequency of these behaviors performed by each student during the exam was extracted, and k-Means clustering was employed for the analysis. The findings revealed that students were classified into six clusters based on their suspicious behaviors. It was found that students in four of the six clusters, representing approximately 33% of the total sample, exhibited suspicious behaviors at varying levels. When the exam scores of these students were compared, it was observed that those who engaged in suspicious behaviors scored, on average, 30–40 points higher than those who did not. Although further research is necessary to validate these findings, this preliminary study provides significant insights into the detection of AI-assisted cheating in online exams using behavior analytics.

## 1. INTRODUCTION

Large language models (LLMs) based on generative artificial intelligence have demonstrated high performance across various domains in benchmark studies (Ali and Xie, 2025). While these capabilities may present significant potential for educational purposes, they also pose substantial risks to assessment, which is a critical component of the educational process. Recent studies have highlighted that students frequently use AI tools such as ChatGPT for cheating purposes (Kazley et al., 2025). Furthermore, specialized AI tools have been developed specifically to facilitate cheating in online exams. These tools operate as browser extensions and are capable of rapidly detecting questions either from text or screenshots, generating answers within seconds. This represents a significant threat to the security of online assessments. However, the extent to which these tools are used by students remains unknown.

It is observed that most studies on the detection of cheating in online exams primarily focus on proctoring and identifying cheating based on students' physical behaviors (Wang et al., 2025, Leong, 2025). On the other hand, research addressing AI-assisted cheating has largely concentrated either on detecting AI-assisted plagiarism (Steponenaite and Barakat, 2023) or on pedagogical strategies aimed at preventing AI-assisted cheating (Leaton Gray et al., 2025). The commonality among these studies is the recognition that generative AI technologies, such as large language models, constitute a significant threat to academic integrity.

It is noteworthy that there are very few studies aiming to detect AI-assisted cheating in online exams based on students' exam-taking behaviors. On the other hand, previous studies have shown that students' exam logs provide valuable insights into their exam-taking behaviors (Yıldırım et al., 2023). Therefore, this study investigates whether it is possible to identify behaviors related to AI-assisted cheating by analyzing students' online exam-taking behaviors.

## 2. METHOD

In this study, exam data from 52 students were analyzed. The data were collected from first-year undergraduate students enrolled in the Department of Computer Education and Instructional Technology during the final exam of the Introductory Programming course. The students took the exam using a web-based exam tool developed by the author. The exam consisted of 25 multiple-choice questions. To enhance exam security, both the order of the questions and the order of the options were randomized for each student. Questions were presented one at a time on the students' screens, and navigation between questions was not permitted. Additionally, the time allowed for each question was limited to 90 seconds. When the allotted time expired, the system was configured to automatically proceed to the next question. The exam was administered in a computer laboratory under the supervision of the course instructor. Students participated in the exam either using computers available in the laboratory or their personal computers. Prior to the exam, students were informed about the academic integrity policy regarding cheating. Students were instructed not to navigate to any page or application other than the exam page during the exam.

In addition to student responses, the developed exam system also recorded students' interactions while answering questions. These interactions included mouse clicks, text selection, loss of page focus, copying, pasting, and so on. Within the scope of this study, three metrics that could serve as indicators of AI-assisted cheating were utilized: the number of text selections (TSC), the number of right-click actions (RCC), and the number of times the page focus was lost (FLC). These metrics were determined by examining popular browser extensions commonly used for cheating, such as AnswerAI.

During the data preprocessing stage, these metrics were calculated for each question individually, and then aggregated totals were obtained for the entire exam. Subsequently, the data were analyzed using clustering method. Clustering analysis is an unsupervised machine learning approach employed to group similar patterns within the data (Demšar and Zupan, 2024). The k-Means clustering algorithm was used for this purpose. All analyses were conducted using the Orange Data Mining software (Demšar et al., 2013). The optimal number of clusters was then determined by evaluating silhouette scores (Rousseeuw, 1987) across a range of 2 to 8 clusters. In other words, the data mining tool tested different cluster solutions on the normalized data and selected the number of clusters that yielded the highest silhouette score as the optimal solution.

## 3. RESULTS

As a result of the clustering analysis, the optimal number of clusters was determined to be six, and students were subsequently grouped into six distinct clusters based on their suspicious behaviors during the exam. Figure 1 presents a graph displaying the cluster means for the variables used in the clustering process. The distribution of students across the clusters is as follows: Cluster 1 (C1) included 3 students, Cluster 2 (C2) included 34 students, Cluster 3 (C3) included 6 students, Cluster 4 (C4) included 4 students, Cluster 5 (C5) included 1 student, and Cluster 6 (C6) included 4 students.

An examination of the cluster means presented in Figure 1 indicates that students were divided into distinct groups based on the type and frequency of suspicious behaviors exhibited during the exam. Students in Clusters 1 (C1), 4 (C4), and 6 (C6) demonstrated similar behavioral patterns, characterized by frequent occurrences of text selection, focus loss, and right-click actions—all considered suspicious behaviors. In contrast, students in Cluster 2 (C2) did not exhibit any suspicious behaviors. Students in Cluster 3 (C3) displayed a low frequency of text selection, did not use right-click actions, but frequently lost focus on the exam page. The single student in Cluster 5 (C5) exclusively engaged in a high number of text selections, without exhibiting right-click or focus loss behaviors.

Figure 1. Cluster means[*]

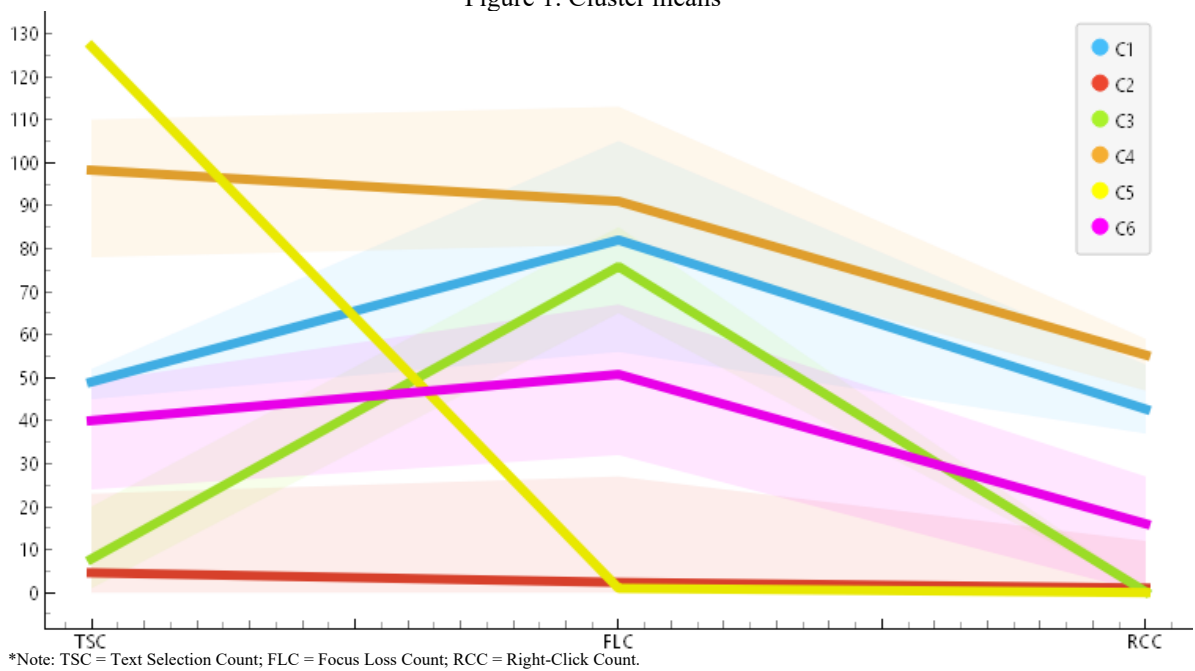

*Note: TSC = Text Selection Count; FLC = Focus Loss Count; RCC = Right-Click Count.

Descriptive statistics regarding the exam scores of students in each cluster were presented in Table 1. According to these results, students in Cluster 2 (C2), who did not exhibit any suspicious behaviors, achieved an average score of 40.94. In contrast, students in Clusters 1 (C1), 4 (C4), and 6 (C6), who demonstrated high frequencies of suspicious behaviors, obtained average scores of 82.67, 77.27, and 57.00, respectively. Students in Cluster 3 (C3), who only showed a high frequency of focus loss, had an average score of 72.00, while the single student in Cluster 5 (C5), who exhibited only a high number of text selection actions, scored 32.00.

Table 2. Descriptive statistics for exam scores

| Cluster | N | Mean | Std. | Median |
|---------|-----|-------|-------|--------|
| C1 | 3 | 82.67 | 7.54 | 88.00 |
| C2 | 34 | 40.94 | 18.36 | 42.00 |
| C3 | 6 | 72.00 | 10.83 | 70.00 |
| C4 | 4 | 77.27 | 4.95 | 76.54 |
| C5 | 1 | 32.00 | 0.00 | 32.00 |
| C6 | 4 | 57.00 | 9.11 | 54.00 |

## 4. DISCUSSION AND CONCLUSION

AI cheating extensions that can be installed on web browsers typically operate as follows: first, the user selects the question text; next, they right-click to open the context menu and choose the extension from the menu. The extension then pops up a window on the right side of the page, providing an answer along with some explanations. Another method of cheating, without using a browser extension, involves copying or taking a screenshot of the question and submitting it to AI tools such as ChatGPT or Gemini. In order to use such cheating methods, students must engage in behaviors such as text selection, right-clicking, and loss of page focus when using browser extensions, or copying and taking screenshots when submitting questions to AI tools. In standard exam-taking behavior, there is no situation that would require the constant, simultaneous use of all three behaviors. However, students in Clusters 1 (C1), 4 (C4), and 6 (C6) (N = 11) demonstrated all three of these suspicious behaviors, though at varying levels. The exam scores of students in these clusters were also substantially higher than those of students in Cluster 2 (C2), who did not exhibit any suspicious behaviors. Therefore, there is a strong likelihood that students in C1, C4, and C6 made use of such AI extensions.

For Cluster 3 (C3), the six students exhibited high rates of focus loss, but did not demonstrate text selection or right-click behaviors. Frequent focus loss indicates that these students were often switching between the exam screen and other tabs or applications during the exam. Such behavior is also inconsistent with standard exam-taking practices, particularly as students were explicitly instructed prior to the exam not to navigate away from the exam page. Notably, the average exam scores of students in this cluster were approximately 35 points higher than those in the non-suspicious group (C2). This raises suspicions that these students may have used AI tools such as ChatGPT in another tab or window during the exam.

On the other hand, the student in Cluster 5 (C5) performed a high number of text selections, but did not engage in right-clicking or focus loss behaviors. This student's exam score was also significantly below the average. It was observed during the exam that some students tend to select text while reading the questions, employing it as a reading strategy. Therefore, text selection alone is not necessarily indicative of cheating.

The likelihood that students cheated using sources other than AI is quite low, as the structure of the exam system presented all questions and answer options to each student in a random order. Students were prevented from returning to previous questions, and the time they could spend on each question was limited. Therefore, the possibility of students copying from one another is significantly reduced. As a result, although these results alone may not be sufficient to conclusively prove cheating, the findings clearly indicate that AI-based cheating constitutes a significant threat to the integrity of online exams. Even in a proctored, face-to-face exam setting, 32.69% of students exhibited suspicious behaviors potentially related to AI-assisted cheating. This study also contributes to future research by identifying behavioral indicators that may be associated with AI-assisted cheating.